\documentclass[11pt,a4paper]{article}
\pdfoutput=1
\usepackage{jheppub}
\usepackage{graphicx}
\usepackage{amsmath}
\usepackage{amsfonts}
\usepackage{mathrsfs}
\usepackage{amssymb}
\usepackage{subfigure}
\usepackage{multicol}
\usepackage{multirow}
\usepackage{array}
\usepackage{slashed}
\usepackage{units}
\usepackage{bbm}
\usepackage{framed}
\usepackage[normalem]{ulem}
\usepackage{mathtools}
\renewcommand{\S}{\mathcal{S}}

\renewcommand{\O}{\mathcal{O}}

\newcommand{\Z}{\mathbb{Z}}
\newcommand{\keV}{{\rm keV}}
\newcommand{\MeV}{{\rm MeV}}
\newcommand{\GeV}{{\rm GeV}}
\newcommand{\TeV}{{\rm TeV}}

\newcommand{\vev}[1]{\langle #1 \rangle}

\newcommand{\blue}[1]{\color{blue} #1 \color{black}}

\title{\blue{On the Viability of Minimal Neutrinophilic Two-Higgs-Doublet Models}}
\author[a]{P. A. N. Machado,}
\author[b]{Y. F. Perez,} 
\author[b,c]{O. Sumensari,} 
\author[d,e]{Z. Tabrizi,}
\author[b]{and R. Zukanovich Funchal}

\emailAdd{pedro.machado@uam.es}
\emailAdd{yfperezg@if.usp.br}
\emailAdd{olcyr.sumensari@usp.br}
\emailAdd{tabrizi.physics@ipm.ir}
\emailAdd{zukanov@if.usp.br}

\affiliation[a]{Departamento de F\'{\i}sica Te\'orica and Instituto de F\'{\i}sica Te\'orica, 
IFT-UAM/CSIC,\\ Universidad Aut\'onoma de Madrid, Cantoblanco, 28049, Madrid, Spain}

\affiliation[b]{Departamento de F\'{\i}sica Matem\'atica, Instituto de F\'{\i}sica, Universidade de S\~ao
  Paulo, \\ C.\ P.\ 66.318, 05315-970 S\~ao Paulo, Brazil}

\affiliation[c]{Laboratoire de Physique Th\'eorique (B\^at.210), \\
Universit\'e Paris Sud and CNRS (UMR 8627), F-91405 Orsay-Cedex, France}

\affiliation[d]{School of Particles and Accelerators,\\
Institute for Research in Fundamental Sciences (IPM),
 P.O.Box 19395-1795, Tehran, Iran}

\affiliation[e]{Instituto de F\'{\i}sica Gleb Wataghin, Universidade
  Estadual de Campinas (UNICAMP), Rua S\'{e}rgio Buarque de Holanda,
  777, Campinas, SP, 13083-859, Brazil}

\abstract{We study the constraints that electroweak precision data can impose, after the discovery of the Higgs boson by the LHC, on neutrinophilic two-Higgs-doublet models which comprise
  one extra $SU(2)\times U(1)$ doublet and a new symmetry, namely
  a spontaneously broken $\Z_2$ or a softly broken global $U(1)$.  In
  these models the extra Higgs doublet, via its very small vacuum
  expectation value, is the sole responsible for neutrino masses. We
  find that the model with a $\Z_2$ symmetry is basically ruled out by
  electroweak precision data, even if the model is slightly extended to include
  extra right-handed neutrinos, due to the presence of a very light
  scalar.  While the other model is still perfectly viable, the
  parameter space is considerably constrained by current data,
  specially by the $T$ parameter. In particular, the new charged and neutral scalars must have very similar masses.}


\begin{document}
{\flushleft{\blue{FTUAM-15-15}}\quad{\blue{IFT-UAM/CSIC-15-057} 
\quad\blue{LPT-Orsay-15-58}\\}}
\maketitle

\section{Introduction}
\label{sec:intro}
The smallness of neutrino masses suggests a mass generating mechanism
distinct from the usual Higgs mechanism, which resides in a scale
different from the electroweak one. From neutrino oscillation
experiments, we know that neutrinos are massive and that mass and
flavor eigenstates do not coincide. Besides, other
terrestrial~\cite{Aseev:2011dq, Kraus:2004zw} and
cosmological~\cite{Hinshaw:2012aka, Ade:2015xua} experiments indicate
that neutrino masses should be below the eV scale. Therefore, if the
same Higgs mechanism is responsible for the top and neutrino masses,
then the Yukawa couplings would span twelve orders of magnitude,
evincing an unpleasant and inexplicable hierarchy. 

A well known alternative is the seesaw
mechanism~\cite{Minkowski:1977sc, Mohapatra:1979ia, Schechter:1980gr}.
In this scenario, the light neutrino masses are suppressed by some
heavy physics, for instance, right-handed Majorana neutrino 
masses~\cite{Minkowski:1977sc, Mohapatra:1979ia, Yanagida:1979as,
  GellMann:1980vs}. What typically happens is that the scale at which
new physics can be found is extremely high, much above the TeV scale,
rendering the model intangible, except for the possible presence of
neutrinoless double beta decay\footnote{Nevertheless, there are
  alternative models which exhibit a low scale, as for instance the
  inverse seesaw scenario~\cite{Mohapatra:1986aw, Mohapatra:1986bd,
    Bernabeu:1987gr}.}. The latter could also originate from some
physics that do not comprise the main contribution to neutrino
masses~\cite{Schechter:1981bd, Duerr:2011zd}, and hence it does not
consist of a test of the seesaw mechanism by itself.

Another possibility is to generate neutrino masses by a copy of the
Higgs mechanism, having a second Higgs doublet, but with a much
smaller vacuum expectation value (vev). This can be achieved in a two-Higgs-doublet model (2HDM) where one of the scalars gives mass to the
charged fermions, while the other one acquires a very small vev and
generates neutrino masses with $\O(1)$ Yukawa couplings, a
\emph{neutrinophilic} 2HDM. As a consequence, neutrino masses would
generically require new physics at the TeV scale (or even lower).  For
instance, by imposing a lepton number  symmetry and adding three
right-handed neutrinos which carry no lepton number, a type I seesaw
mechanism can be realized below the TeV
scale~\cite{Ma:2000cc}. Moreover, lepton number could be conserved and
a $\Z_2$ symmetry~\cite{Gabriel:2006ns, Haba:2011nb} or a global
$U(1)$~\cite{Davidson:2009ha} could be used to prevent the SM Higgs
boson to couple to neutrinos, yielding Dirac neutrinos. Also, the 2HDM
could be augmented by a type III seesaw and a $\mu-\tau$ symmetry,
giving rise to interesting LHC
phenomenology~\cite{Bandyopadhyay:2009xa}; or by a singlet scalar and
a $\Z_3$ symmetry, possibly generating lepton flavor
violating signals~\cite{Haba:2010zi}. It is important to note that such models are stable against radiative
corrections~\cite{Morozumi:2011zu, Haba:2011fn}.

On general grounds, a new symmetry is typically invoked to prevent the
first scalar doublet from coupling to neutrinos as well as to enforce
the second one to interact only with them.  These models introduce a
minimal new field content which should materialize as particles below
the TeV scale.
The presence of such a low scale in the theory might have important
phenomenological consequences, like the presence of light scalar
particles (for instance, supernova energy loss strongly constrains such
scenarios~\cite{Zhou:2011rc}). After the discovery of a 125 GeV scalar
by the LHC experiments, new limits from electroweak precision data can
be derived on the allowed parameter space of such models.  The purpose
of this manuscript is to investigate to what extent these minimal
neutrinophilic 2HDMs can survive electroweak precision data
scrutiny.

In sec.~\ref{sec:2hdm} we briefly review the neutrinophilic 2HDMs
which we will study in this work. In sec.~\ref{sec:constraints} we
describe the theoretical and experimental constraints that will be
imposed on these models in sec.~\ref{sec:ana}. Finally, in
sec.~\ref{sec:conc} we present our conclusions.

\section{Neutrinophilic Two-Higgs-Doublet Models}
\label{sec:2hdm}
We first start by making general considerations on the 2HDM and the link to neutrino masses.  The most general scalar
potential for a 2HDM is
\begin{align}
\label{v2hdm}
V(\Phi_1,\Phi_2)&=m_{11}^2\Phi_1^\dagger\Phi_1 +m_{22}^2\Phi_2^\dagger\Phi_2-(m_{12}^2\Phi_1^\dagger\Phi_2+\text{h.c.})\nonumber\\
&+\dfrac{\lambda_1}{2}(\Phi_1^\dagger \Phi_1)^2+\dfrac{\lambda_2}{2}(\Phi_2^\dagger \Phi_2)^2
+\lambda_3 \Phi_1^\dagger\Phi_1 \Phi_2^\dagger\Phi_2+\lambda_4 \Phi_1^\dagger\Phi_2 \Phi_2^\dagger\Phi_1\\
&+\left[\dfrac{\lambda_5}{2}(\Phi_1^\dagger\Phi_2)^2 +\left(\lambda_6 \Phi_1^\dagger\Phi_1 + \lambda_7 \Phi_2^\dagger\Phi_2\right)\Phi_1^\dagger\Phi_2 + \text{h.c.}\right],\nonumber
\end{align}
where $\Phi_1$ and $\Phi_2$ are two scalar doublets with hypercharge
$Y=+1$. For the vacuum expectation values of the two scalars, we adopt
the notation $\vev{\Phi_1}=v_1/\sqrt{2}$, $\vev{\Phi_2}=v_2/\sqrt{2}$,
and we pick $\Phi_2$ to be the one responsible for neutrino masses. In
order to have sizable Yukawa coupling for neutrinos, it is required
that $v_2 \ll v_1 \sim 246~\GeV = v$, where $v^2=v_1^2+v_2^2$.  In
principle, the parameters $m_{12}^2$, $\lambda_5$, $\lambda_6$, and
$\lambda_7$ can be complex. Nevertheless, in all models we analyze,
the symmetries will forbid both $\lambda_6$ and $\lambda_7$, and only
$m_{12}^2$ or $\lambda_5$ will be allowed to be non-zero. A single
phase of the aforementioned parameters can always be absorbed in a
redefinition of the scalar fields, and therefore we can take all
scalar potential parameters to be real without loss of generality.

To forbid the coupling between neutrinos and $\Phi_1$, a symmetry is
called for.  In this minimal setup, there are two straightforward
examples. The first possibility is a $\Z_2$ symmetry under which only
$\Phi_2$ and the right-handed neutrinos are charged, forcing
$m_{12}=\lambda_6=\lambda_7=0$.  An alternative is
to trade the $\Z_2$ by a global $U(1)$, yielding, in principle,
$m_{12}=\lambda_5=\lambda_6=\lambda_7=0$. In this case, to avoid the
presence of a massless Goldstone boson, a soft breaking is introduced
by having a non-zero but small $m_{12}$. On the other hand, if a
softly broken $\Z_N$, $N>2$, symmetry is postulated, $\lambda_5$ might 
be forcefully zero as well, making this case identical to the $U(1)$ 
scenario. Therefore, the phenomenology of a softly broken $\Z_{N>2}$ model 
is identical to the softly broken $U(1)$ case.  Anyhow, in all realizations
we will study here $\lambda_6=\lambda_7=0$, so these couplings will be
disregarded henceforth.

One last option that one could consider would be to gauge the $U(1)$
symmetry, avoiding the massless Goldstone boson. Nevertheless, in such
a scenario, the corresponding gauge boson as well as one of the
neutral scalars would be extremely light, with mass around the $v_2$
scale. This seems, at first glance, phenomenologically quite problematic. We do not
investigate this possibility here as it would require a completely
different study compared to the other two cases.
 
The two complex scalar $SU(2)$ doublets can be written as
\begin{equation}
\label{Higgs}
\Phi_a 	=
	\begin{pmatrix}
		\phi_a^+\\ (v_a+\rho_a+i\eta_a)/\sqrt{2} 
	\end{pmatrix},\qquad a=1,2.
\end{equation}
After electroweak symmetry breaking, three Goldstone bosons
become the longitudinal modes of the $W$ and $Z$ bosons. Then, the
remaining scalar spectrum is composed of two charged particles,
$H^\pm$, two CP-even neutral bosons, $h$ and $H$, and one CP-odd
neutral boson, $A$. The physical fields are given by
\begin{align}
H^+ =\phi_1^+\sin\beta -\phi_2^+\cos\beta, \qquad \qquad A =\eta_1\sin\beta-\eta_2\cos\beta,\\
h \;\;=-\rho_1\cos\alpha -\rho_2\sin\alpha, \qquad \qquad H =\rho_1\sin\alpha -\rho_2\cos\alpha,
\end{align}
where the angles $\alpha$ and $\beta$ are associated with the rotations
that diagonalize the mass matrices
\begin{align}\label{eq:alpha}
	\tan(2\alpha)&=\dfrac{2(-m_{12}^2+\lambda_{345}\; v_1 v_2)}{m_{12}^2(v_2/v_1-v_1/v_2)+\lambda_1 v_1^2-\lambda_2 v_2^2},\\\label{eq:beta}
	\tan\beta &= \dfrac{v_2}{v_1},
\end{align}
where $\lambda_{345}\equiv\lambda_3+\lambda_4+\lambda_5$. We will see
below that both $\alpha$ and $\beta$ are expected to be very
small. Hence, $h$ behaves very similarly to the SM Higgs, while the scalars $H,A,H^\pm$ develop neutrinophilic interactions in the Yukawa sector, as described below

\begin{equation}
\label{yuk}
\mathcal{L}_{Y}= \frac{m_{\nu_i}}{v_2}H \bar{\nu_i}\nu_i- i \frac{m_{\nu_i}}{v_2}A \bar{\nu_i}\gamma_5\nu_i-\frac{\sqrt{2}m_{\nu_i}}{v_2} [U_{\ell i}^\ast H^+ \bar{\nu_i}P_L \ell+ \mathrm{h.c.}],
\end{equation}
where $m_{\nu_i}$ are neutrino masses and $U_{\ell i}$ is the PMNS matrix. As we will see in sec.~\ref{sec:ana}, the tree-level stationary
conditions on the potential, $\partial V/\partial \Phi_i=0$, can be
used to write the diagonal mass parameters $m_{ii}$ as functions of
$m_{12}^2$, the quartic couplings and the vevs. With that in mind, we
can consider the quartic couplings as free parameters and express them
in terms of the physical masses, vevs and mixing
angles~\cite{Kanemura:2004mg} (see appendix \ref{ap:quartic}).

Next we describe the two specific realizations of the neutrinophilic
scenarios that will be studied in this paper.

\subsection{Neutrinophilic 2HDM: $\Z_2$ symmetry}
\label{subsec:z2}
The model to be studied was proposed by Gabriel and
Nandi~\cite{Gabriel:2006ns}\footnote{The same model was previously
  also discussed in ref.~\cite{Wang:2006jy} where the focus was on the
  origin of the second doublet from neutrino condensation.}.  It
consists of a 2HDM where both the right-handed neutrinos and one of
the scalar doublets, $\Phi_2$, are charged under a $\Z_2$
symmetry. The consequence is that the masses of the charged fermions
come solely from the $\Phi_1$ vev, and neutrinos, which are Dirac
fermions in this scenario as the authors impose lepton number
conservation, couple exclusively to $\Phi_2$.  This extra symmetry
can, in principle, be dropped allowing for Majorana neutrinos with a
low scale realization of the seesaw mechanism.  We will also investigate this
possibility in our analysis.

In the scalar potential (\ref{v2hdm}) of this model, the parameters
$m_{12}^2$ and $\lambda_{6,7}$ will vanish due to the $\Z_2$ symmetry. The
smallness of neutrino masses is explained by the very low scale at which
$\Z_2$ is broken, preferably $v_2\lesssim\O({\rm eV})$.\footnote{
  It is known that the breaking of discrete symmetries leads 
    to the formation of domain walls, which may store unacceptably 
    large quantities of energy, unless the vev responsible for this 
    breaking is below $\mathcal{O}(10^{-2})$ GeV \cite{Kibble:1976sj,Dvali:1994wv}. 
    Nevertheless, as the second scalar has to have a vev small enough 
    to explain neutrino masses, domain walls do not pose a bound 
    on neutrinophilic 2HDMs.} A tiny
$v_2/v_1$ ratio and the absence of an explicit breaking $m_{12}^2$
term leads to almost no mixing between the doublets. The smallness of
$\tan\beta$ and $\tan\alpha$ can be seen from eqs.~(\ref{eq:alpha})
and (\ref{eq:beta}) after imposing $v_2/v_1\rightarrow0$. Therefore,
apart from its couplings to neutrinos, $\Phi_1$ behaves almost
identically to the SM Higgs doublet, so we do not expect any
observable deviation from the Higgs couplings to the SM particles,
except possibly the loop induced couplings, e.g. $h\gamma\gamma$.

The second doublet displays some interesting features. Through the
Yukawa coupling, the neutral components couple almost only to
neutrinos, while the charged scalars mediate interactions between neutrinos and charged leptons (see eq.~(\ref{yuk})). The Yukawas are
ideally expected to be of $\O(1)$. The neutral scalars couple to the
$W$ and $Z$ bosons, but notice that triple gauge couplings (TGCs)
involving only one scalar are highly suppressed by the small vev, $v_2$. 
Obviously, TGCs with two scalars and one gauge boson are present
and may provide a sizeable pair production cross section at colliders,
for instance $pp\to A^* \to H^+H^-$ at the LHC.

The scalar spectrum of this model is quite constrained. By setting
$m_{12}^2=0$ in eqs.~(\ref{eq:mH})-(\ref{eq:mA}), as well as
$\sin^2\alpha,\sin^2\beta\ll 1$, we notice that: (i) $h$ is identified
as the 125~GeV Higgs particle found at the LHC, and this essentially
fixes $\lambda_1 \approx 0.26$ (see eq.~(\ref{lambda1})); (ii) the
neutrinophilic neutral scalar $H$ is extremely light, $m_H
\sim\O(v_2)\ll v$; and (iii) for not so large values of the quartic
couplings, the charged scalars and the pseudoscalar masses are bounded
to be about or below the TeV scale.
When we analyse the viability of this model in sec.~\ref{sec:ana}, it
will turn out that oblique parameters will play a decisive role in
constraining it, due to the peculiar structure of the scalar spectrum.
The sensitivity of the $S$ parameter to the presence of a very light
neutral scalar, $m_H\sim\O(v_2)$, will essentially rule out the model.

\subsection{Neutrinophilic 2HDM: softly broken global $U(1)$ symmetry}
\label{sec:u1}
The second model we study was proposed by Davidson and
Logan~\cite{Davidson:2009ha}.  Analogously to the other scenario, both
$\Phi_2$ and right-handed neutrinos are charged under a new global
$U(1)$. The model spans $\lambda_{5,6,7}=0$ and a small $m_{12}^2$
which breaks the symmetry softly and generates neutrino masses. The
presence of the soft breaking mass term, is required in order to avoid
a massless Goldstone boson which might create problems with cosmology
and electroweak precision data.  Neutrinos are Dirac particles, as the
Majorana mass term is strictly forbidden by the new $U(1)$. From
eq.~(\ref{eq:mA}), we write
\begin{equation}
  m_{12}^2 = \sin\beta\cos\beta \; m_A^2,
\end{equation}
and we observe that to obtain simultaneously $v_2\sim$~eV and $m_A
\sim \mathcal{O}(100~\GeV)$ one would need $m_{12}^2\sim
(200~\keV)^2$.  As said before, to avoid the issues of having a
massless Goldstone, instead of softly breaking the new $U(1)$
symmetry, one could also envisage to gauge it. Nonetheless, the theory
would contain a very light vector resonance as a consequence of the
small vev, and it is not clear if such a model can satisfy all
neutrino data and astrophysical constraints. We do not explore this
possibility here.

The presence of a non-zero $m_{12}^2$ term makes this case fairly
different from the last one. From eq.~(\ref{eq:mh}), we notice that
the mass of the neutrinophilic scalar, $m_H$, increases with $M$, and therefore the $H$ mass in this scenario is not bounded by
$v_2$ as in the previous case.  As we will see later, this will ease
the constraints from the oblique parameters. Combining
eq.~(\ref{lambda2}) with the definition
$M^2=m_{12}^2/(\sin\beta\cos\beta)$, and imposing $\tan\beta=v_2/v\ll 1$, we obtain
\begin{align}
\label{eq:lambda2}
\lambda_2  &=\frac{1}{v^2}\left(-\cot^2\beta M^2+\frac{\cos^2\alpha}{\sin^2\beta} m_H^2+\frac{\sin^2\alpha}{\sin^2\beta}m_h^2\right) \simeq \frac{1}{v_2^2}\left(m_H^2-m_{12}^2 \frac{v}{v_2}\right)+\frac{\sin^2\alpha}{\sin^2\beta}\frac{m_h^2}{v^2},
\end{align}
which indicates that
\begin{equation}
  |m_H^2-m_{12}^2 v/v_2|\lesssim \O(v_2^2).
\end{equation}
To grasp the impact of this conclusion, assume that $m_{12}^2 = m_H^2
v_2/v$. Hence, from eq.~(\ref{eq:mA}) we see that $m_A\approx m_H$, so
the neutrinophilic CP-odd and CP-even scalars, $A$ and $H$, are 
degenerate in mass. We emphasize that this degeneracy by itself is not a fine
tuning of the model: the degenerate spectrum arises naturally given
the symmetries of the scalar potential and the hierarchy between the
vevs. As a last comment, we emphasize that since $m_{12}^2$ is the
only source of $U(1)$ breaking, it is natural in the t'Hooft sense --
$m_{12}^2$ only receives radiative corrections proportional to
itself~\cite{Morozumi:2011zu, Haba:2011fn}.

\section{Theoretical and Experimental Electroweak Data Constraints}
\label{sec:constraints}

\subsection{Theoretical Constraints}

There are a number of conditions to be fulfilled by the scalar
potential. These will be used to constrain the parameter space,
ultimately restricting the range of physical scalar masses, having an
important impact on the phenomenology of the models.  To have
stability at tree-level, the following constraints should be
fulfilled~\cite{hep-ph/0207010}
\begin{equation}
\label{contrainte}
\lambda_{1,2}>0,\qquad{}\lambda_3>-(\lambda_1\lambda_2)^{1/2}\qquad{}\text{and}\qquad{}\lambda_3+\lambda_4-|\lambda_5|>-(\lambda_1 \lambda_2)^{1/2}.
\end{equation}
In addition, the stationary conditions $\partial V/\partial \Phi_i=0$ read
\begin{equation}
\begin{aligned}\label{eq:stationary}
	&\dfrac{\lambda_1}{2}v_1^3+\dfrac{\lambda_{345}}{2}v_1v_2^2+
m_{11}^2v_1-m_{12}^2v_2=0,\\
	&\dfrac{\lambda_2}{2}v_2^3+\dfrac{\lambda_{345}}{2}v_2v_1^2+
m_{22}^2v_2-m_{12}^2v_1=0,
\end{aligned}
\end{equation}
which allow us to write $m_{ii}^2$ as functions of $m_{12}^2$, $v_1$ and $v_2$.
If $m_{12}^2=0$, it is easy to see that there are at least two
equivalent stable solutions, $(v,0)$ or $(0,v)$ (although they may not
be the global minima).
In this case, the vev is precisely the electroweak scale, one of the
scalars is exactly the Higgs and the other one is inert.  For
$m_{12}^2 \neq 0$, these equations cannot be solved
analytically. Nevertheless, if $m_{12}^2\ll v^2$ a perturbative
approach yields
\begin{equation}
   v_1\approx v, \qquad v_2\approx\dfrac{m_{12}^2}{\frac{\lambda_{345}}{2}\, v^2+m_{22}^2}v,
\end{equation}
and a symmetric solution interchanging the indices $1\leftrightarrow
2$, which reveals that the small vev necessary to satisfactorily
explain small neutrino masses might require a correspondingly small
$m_{12}^2$ parameter. This can be understood intuitively, as the
breaking of the $U(1)$ happens only through the soft breaking term
$m_{12}^2$.  In general, there can be more than one solution
satisfying the stationary conditions~(\ref{eq:stationary}), and hence
different non-trivial and non-degenerate minima $(v_1,v_2)$ and
$(v_1',v_2')$ might coexist. It is possible to check analytically if
the chosen vacuum is the deepest one in the potential for a 2HDM with
$\lambda_{6,7}=0$~\cite{1303.5098}. In this case, the potential
describes a $\Z_2$ symmetry softly broken by $m_{12}^2$. Both models
we deal with here are special cases of such scenario. In the absence of an
explicit breaking, that is $m_{12}^2=0$, there can be multiple minima,
but they are degenerate and hence stability is not threatened. This is
the case of the $\Z_2$ model we analyze. For the softly broken $U(1)$
model, it can be shown that the chosen vacuum is the deepest one (at
tree-level) if and only if the following condition is satisfied \cite{1303.5098}:

\begin{equation}
	D=m_{12}^2(m_{11}^2-\kappa^2 m_{22}^2)(\tan\beta-\kappa)>0,
\end{equation}
with $\kappa=\sqrt[4]{\lambda_1/\lambda_2}$. Although for a general
2HDM scenario this bound may be important, for the neutrinophilic case
we have checked that it does not lead to any significant effect on the
parameter space, after the other constraints are taken into account,
but we include it in the analysis of the softly broken $U(1)$ model
for completeness.

Another theoretical requirement is to satisfy the tree-level pertubative
unitarity condition~\cite{Kanemura:1993hm, Arhrib:2000is, hep-ph/0508020}.
If the quartic couplings are too large, the lowest order amplitudes
for scalar--scalar scattering may violate unitarity at high enough
scales, requiring additional physics to mitigate this issue. To obtain
the constraint, the scalar--scalar $S$ matrix is computed and the
following conditions are imposed on its eigenvalues
\begin{equation}\label{eq:unitarity}
	|a_{\pm}|,|b_{\pm}|,|c_{\pm}|,|f_{\pm}|,|e_{1,2}|,|f_1|,|p_1|< 8 \pi,
\end{equation}
where 
\begin{subequations}
\begin{alignat}{9}
a_\pm &= \frac{3}{2}(\lambda_1+\lambda_2)\pm \sqrt{\frac{9}{4}(\lambda_1-\lambda_2)^2+(2\lambda_3+\lambda_4)^2}, \\
b_\pm &= \frac{1}{2}(\lambda_1+\lambda_2)\pm \frac{1}{2}\sqrt{(\lambda_1-\lambda_2)^2+4\lambda_4^2}, \\
c_\pm &= \frac{1}{2}(\lambda_1+\lambda_2)\pm \frac{1}{2}\sqrt{(\lambda_1-\lambda_2)^2+4\lambda_5^2}, \\
f_+ &= \lambda_3 + 2\lambda_4 + 3 \lambda_5, \\
f_- &= \lambda_3 + \lambda_5, \\
e_1 &= \lambda_3 + 2\lambda_4 - 3 \lambda_5, \\
e_2 &= \lambda_3- \lambda_5, \\
f_1 &= \lambda_3 + \lambda_4, \\
p_1 &= \lambda_3 - \lambda_4. 
\end{alignat}
\end{subequations}
To have an idea of the impact of these bounds, one can conservatively
assume that all $|\lambda_i|$ should be smaller than $8\pi$ (the
actual bound is always more stringent than that). Some authors
prefer to use a stronger limit of $4\pi$. We checked that this does not
change very much the allowed regions.

Evidently, even if tree-level unitarity is satisfied, loop corrections
could still play an important role leading to violation of unitarity
at some scale and thus demanding the presence of new physics below
such energies. This could be particularly relevant when some of the
tree-level constraints are just barely satisfied, as the size of the
quartic couplings could enhance the loop contributions. Nevertheless,
we only take into account unitarity constraints at tree-level, as a
full one loop evaluation of the parameter space is beyond the
scope of this manuscript.

\subsection{Electroweak Data Constraints}
\label{sub:ew-cons}
{\bf Oblique Parameters.}  The impact of a second Higgs doublet in the
so-called electroweak precision tests (EWPT), encoded in the
Peskin-Takeuchi parameters $S$, $T$, and $U$~\cite{Peskin:1991sw}, has
been studied in the literature to a great extent (see for instance
refs.~\cite{0711.4022, 0802.4353, 1011.6188}). These are radiative
corrections to the gauge boson two point functions, known as oblique
corrections. For the precise expressions of $S$, $T$, $U$, we point
the reader to the aforementioned references. 

The $S$ parameter encodes the running of the neutral gauge bosons two
point functions ($ZZ$, $Z\gamma$ and $\gamma\gamma$) between zero
momentum and the $Z$ pole. Therefore, it should be specially sensitive
to new physics at low scales, particularly below the $Z$ mass. Thus,
we expect it to be important in the presence of very light neutral
scalars, as is the case for the $\Z_2$ model.  The $T$ parameter
measures the breaking of custodial symmetry at zero momentum, that is,
the difference between the $WW$ and the $ZZ$ two point functions at
$q^2=0$. It usually plays a significant role in constraining the
parameter space of particles charged under $SU(2)_L$. Splitting the
masses of particles in a doublet breaks custodial symmetry and affects
$T$. As we will see later, in the softly broken $U(1)$ scenario, the
$T$ parameter will provide the major constraint on the mass splitting
$m_{H^\pm}-m_A$, forcing the scalar spectrum of this model to be
somewhat degenerate.  Last, and this time least, the $U$ parameter (or
better, the combination $S+U$) is somewhat similar to $S$ but for the
$W$ bosons, being sensitive to light charged particles in the loops.
Given the fact that light charged particles are excluded by LEP
data~\cite{Abbiendi:2013hk, Agashe:2014kda}, usually $U$ is the least
important of these three precision parameters, having a minor impact
on the model phenomenology, we have checked that this is indeed the
case for all scenarios analyzed here.

To evaluate the impact of the EWPT on the neutrinophilic 2HDM
scenarios, we calculate $S$, $T$, and $U$ using the results available
in ref.~\cite{1011.6188}, and we use the latest GFITTER values for the
best fit, uncertainties and covariance matrix~\cite{Baak:2014ora},
\begin{equation}
  \begin{aligned}
    & \Delta S^{SM} = 0.05\pm0.11,\\
    & \Delta T^{SM} = 0.09\pm0.13,\\
    & \Delta U^{SM} = 0.01\pm0.11,\\
  \end{aligned}
\qquad\qquad
V = \left(\begin{array}{ccc}
1 & 0.90 & -0.59\\
0.90 & 1 & -0.83\\
-0.59 & -0.83 & 1
\end{array}\right),
\end{equation}
composing the $\chi^2$ function as
\begin{equation}
  \chi^2= \sum_{i,j}(X_i - X_i^{\rm SM})(\sigma^2)_{ij}^{-1}(X_j - X_j^{\rm SM}),
\end{equation}
with $X_i=\Delta S, \Delta T, \Delta U$ and the covariance matrix
$\sigma^2_{ij}\equiv\sigma_iV_{ij}\sigma_j$, in which
$(\sigma_1,\sigma_2,\sigma_3)=(0.11,0.13,0.11)$. As we are interested
in the goodness of fit of the model to the EWPT data, the 1, 2, and
3$\sigma$ regions are calculated using $\chi^2 = 3.5,8.0,14.2$,
respectively.

{\bf Higgs invisible width.} 
When the first doublet acquires a vev, triple scalar vertices like
$h\S\S$ ($\S=H,A$) are induced. Therefore, light neutral scalars with $2
m_\S < m_h$ could contribute to the Higgs invisible width $h\to \S \S$,
and sequentially $\S\to\bar\nu\nu$. Because of the small $\tan\beta$ of
the model, the Higgs boson couplings to the Standard Model particles is
basically unchanged. Hence, the contribution to the Higgs total width
due to the invisible decay will suppress all Standard Model branching
fractions by the ratio $\Gamma_h^{\rm SM}/\Gamma_h^{\rm new}$. In this
scenario, as the only modification to the Higgs branching fractions is
the addition of an invisible channel, the LHC 8 TeV data bound is
${\rm BR}(h\to {\rm invisible})<0.13$ at 95\% CL~\cite{Ellis:2013lra}.

In our framework, the decay rate of such a process is given
by~\cite{Bernon:2014nxa}
\begin{align}
&\Gamma(h\to \S \S) = \frac{g_{h\S\S}^2}{32 \pi m_h} \sqrt{1-\frac{4 m_\S^2}{m_h^2}}, 
\qquad
\end{align}
with
\begin{align}
&g_{hAA}=\frac{1}{2v}\Bigg{[}(2 m_A^2-m_h^2)\frac{\sin(\alpha-3\beta)}{\sin 2\beta}
\nonumber\\
&\qquad\qquad\qquad +(8m_{12}^2-\sin 2\beta(2m_A^2+3 m_h^2)) 
  \frac{\sin(\beta+\alpha)}{\sin^2 2\beta}\Bigg{]},\label{eq:HAA}\\
&g_{hHH}=-\frac{1}{v} \cos(\beta-\alpha)\Bigg[\frac{2 m_{12}^2}{\sin 2\beta} 
  +\left(2 m_H^2+m_h^2-\frac{6 m_{12}^2}{\sin 2\beta}\right)
  \frac{\sin 2\alpha}{\sin 2\beta}\Bigg].\label{eq:Hhh}
\end{align}
While the couplings between the SM Higgs and the SM fermions,
$g_{hff}=m_f/v$, are well below one due to the suppression by the EW
scale (except for the top, to which the Higgs cannot decay), the
trilinear scalar couplings are typically much larger, $g_{h\S\S}\sim
m_h^2/v \sim 60\, \mathrm{GeV}$, unless there is some sort of
cancellation happening~\cite{Bernon:2014nxa}. Therefore, SM Higgs
decays to lighter scalars may have an important phenomenological
impact, see e.g. ref.~\cite{Seto:2015rma}, specially because
the total Higgs width in the Standard Model is predicted to be very
small, around $4.07~\MeV$~\cite{Denner:2011mq}.

{\bf Higgs to diphoton.} The charged scalars will contribute to the
$h\to\gamma\gamma$ width, and thus we also analyse the impact on this
observable\footnote{The $h\to Z\gamma$ decay will also be modified,
  but due to the smaller branching ratio and subsequent suppression by
  requiring the $Z$ to decay leptonically, we do not expect it to
  provide any significant sensitivity in the near future.}. The $h$
diphoton width is a destructive interference effect mainly between $W$
and top loops, where the latter dominate. Charged scalars contribute
with the same sign as the $W$, and their contribution usually do not
overcome the top one. Therefore, we expect $h\to\gamma\gamma$ to be
somewhat suppressed in most cases.  The expression for the
$h\to\gamma\gamma$ width at one loop can be found in many papers, see,
for instance ref.~\cite{Almeida:2012bq}. For reference, 
the current ATLAS$+$CMS combination value of the Higgs to diphoton signal strength is 
$\mu_{\gamma\gamma}=1.16^{+0.20}_{-0.18}$~\cite{ATLAS-CONF-2015-044}.

{\bf $Z$ invisible width.}  We also have to consider possible extra
contributions to the $Z$ invisible width coming from the decays $Z \to
\S \nu \bar{\nu}$ with $\S=A,H$ and $m_\S<m_Z$. In the model with a softly broken $U(1)$ symmetry, the expression for $\Gamma(Z\to \S \nu\bar{\nu}) = \Gamma(Z\to A \nu\bar{\nu})+\Gamma(Z\to H \nu\bar{\nu})$ can be easily calculated and reads
\begin{align}
\label{eq:zinv}
  \Gamma(Z\to \S \nu\bar{\nu}) &= 
       \dfrac{1}{384 \pi^3 m_Z^5} \left(\frac{g}{2 \cos\theta_W}\right)^2 \frac{m^2_{\nu,\mathrm{tot}}}{v_2^2} \int_0^{(m_Z-m_\S)^2}\mathrm{d}q^2\, \frac{\lambda^{1/2}(q^2,m_Z^2,m_\S^2)}{(q^2-m_\S^2)^2+m_\S^2 \Gamma_\S^2}\notag\\
       &\times\Bigg{[}g_\S(q^2)+\frac{f_\S(q^2)}{\lambda^{1/2}(q^2,m_\S^2,m_Z^2)}\coth^{-1}\left(\frac{m_Z^2+m_\S^2-q^2}{\lambda^{1/2}(q^2,m_Z^2,m_\S^2)}\right) \Bigg{]},
\end{align}
where $m_\S$ is the mass of the neutrinophilic scalars $H$ and $A$, which are degenerate in mass, and the total width is given by

\begin{equation}
\Gamma_\S = \frac{m_\S}{8\pi}\frac{m_{\nu,\mathrm{tot}}^2}{v_2^2}.
\end{equation} 

\noindent We also define $\lambda(a^2,b^2,c^2)=(a^2-(b-c)^2)(a^2-(b+c)^2)$ and
\begin{align}
f_\S(q^2) &= 4 m_Z^2 \left[ (m_\S^2-q^2)(m_\S^4 - m_Z^2 q^2+ q^4+m_\S^2(m_Z^2-4 q^2))+\Gamma_\S^2 m_\S^2 (m_\S^2+m_Z^2-q^2)\right],\\
g_\S(q^2) &= 4 m_\S^4(q^2-m_Z^2)+m_\S^2[4 m_Z^2(2q^2-\Gamma_\S^2)+q^2(\Gamma_\S^2-8 q^2)]+q^2(m_Z^4-8m_Z^2 q^2+4q^4).
\end{align}

\noindent The ratio between $m^2_{\nu,{\rm tot}}\equiv\sum
m_{\nu_i}^2$ and $v^2_2$ arrives from the neutrino Yukawas. Clearly,
if the Yukawas are small, both widths vanish, so we expect this bound
to be more significant for lower $v_2$ and larger neutrino masses.  To
constrain extra contributions from new physics to the Z invisible
width, we use LEP result
$\Gamma^{\rm{exp}}(Z\to\rm{invisible})=499.0(15)$~MeV and the Standard
Model prediction
$\Gamma^{\rm{SM}}(Z\to\rm{invisible})=501.69(6)$~MeV~\cite{Agashe:2014kda},
which yields $\Gamma^{\rm{NP}}(Z\to\rm{invisible})<1.8$ MeV at 3$\sigma$ (notice that there is a mild $2\sigma$ discrepancy between
the data and the SM predicted value). In the case of the
$\mathbb{Z}_2$ symmetry model, one must take care while doing the
computation, since $m_H\ll m_Z$, as the expression for the width has
an infrared divergence, which cancels out with radiative
conditions. As we will see in sec.~\ref{sec:ana}, the other
constraints will exclude most of the parameter space of this
model. For this reason we will not discuss the constraints from the
$Z$ invisible width in this scenario.

{\bf Collider bounds on charged scalars.} The charged scalars can be
pair produced directly at colliders via $s$-channel off shell photon
or $Z$ exchange. Due to the neutrinophilic character of the second
Higgs doublet and small admixture with the SM degrees of freedom, the
charged scalars decay almost only to $\ell\nu$. Therefore, we use the 
corresponding LEP bound, {\em i.e.} $m_{H^\pm}>80~\GeV$~\cite{Abbiendi:2013hk,
  Agashe:2014kda}.

It is not clear how LHC data improves the
situation. There has been some studies on the LHC sensitivity to such
charged scalars, mainly focused on 14~TeV center of mass
energy~\cite{Davidson:2010sf, Morozumi:2012sg, Morozumi:2013rfa,
  Maitra:2014qea, Choi:2014tga}, but to the best of our knowledge, there has been no
dedicated experimental search for charged scalars in neutrinophilic
2HDMs.  As $v_2$ is very small, the main production modes of $H^\pm$
would be pair production through vector boson fusion or off-shell
$s$-channel photon and $Z$ exchange, and the tipical $t \to H^+ b$
would be absent due to small $\tan\beta$.  The LHC sensitivity then
would come mainly from opposite sign dilepton plus missing energy,
which has SM $W$ pair production as an irreducible
background. Moreover, the branching ratios of the charged scalar
depend on the neutrino masses and the mass ordering. If the $\tau\nu$
branching ratio is dominant, the sensitivity is expected to be
smaller. Therefore, to be conservative, we will scan the parameter
space considering only the LEP bound.

{\bf Anomalous magnetic moments and other constraints.} In principle,
the charged scalars could also contribute to charged lepton $g-2$
values, but the corresponding amplitude at one loop is suppressed by
$m_\ell^4/m_{H^\pm}^4$ (see ref.~\cite{1409.3199} for a recent
analysis on the impact of a second Higgs doublet on the muon
$g-2$). We have checked that the 1-loop contribution to both muon and
electron $g-2$ is negligible due to that suppression, while the tau
$g-2$ is not measured with enough precision to pose a bound. For a
general 2HDM, it has been noticed that two loop Barr-Zee
diagrams~\cite{hep-ph/0009292} can be more important than 1-loop
contributions, but this is not the case in the neutrinophilic 2HDM, as
the charged lepton couplings to $H$ and $A$ are suppressed by
$\tan\beta$~\footnote{There would be a small contribution due to
  modifications of the $h\to\gamma\gamma$ coupling, but Higgs data
  already constraint it to the level that there is no observable
  modification to the muon $g-2$.}. Therefore we conclude that the
electron, muon and tau $g-2$ measurements do not pose any bound on
this scenario.

Flavor physics constraints have also been studied in the
literature. The charged scalars will mediate lepton flavor violating
decays. In $\mu\to e\gamma$, for instance, the additional branching
ratio is proportional to $(m_{H^\pm}v_2)^{-4}$.  Because of that, it
is always possible to evade this bound for large enough values of
$v_2$ (or $m_{H^\pm}$). The limits we derive here are independent of a
particular choice of $v_2$ as they concern directly the spectrum.

\section{Analysis of the Models}
\label{sec:ana}
For each model we generate $\approx 10^7$ points. For each of those points 
we calculate the corresponding scalar potential parameters and verify 
if they fulfill the constraints described in sec.~\ref{sec:constraints}. 
We only show on our plots the allowed points, which are about $10\%$ of the generated ones. Unless stated
otherwise, the points are color coded accordingly to the fit to EWPT
data: blue, green, and red correspond to the 1, 2, and 3$\sigma$
allowed regions, while gray points are excluded at $3\sigma$ or more.

\begin{figure}[t]
  \begin{center}
    \includegraphics[scale=0.71]{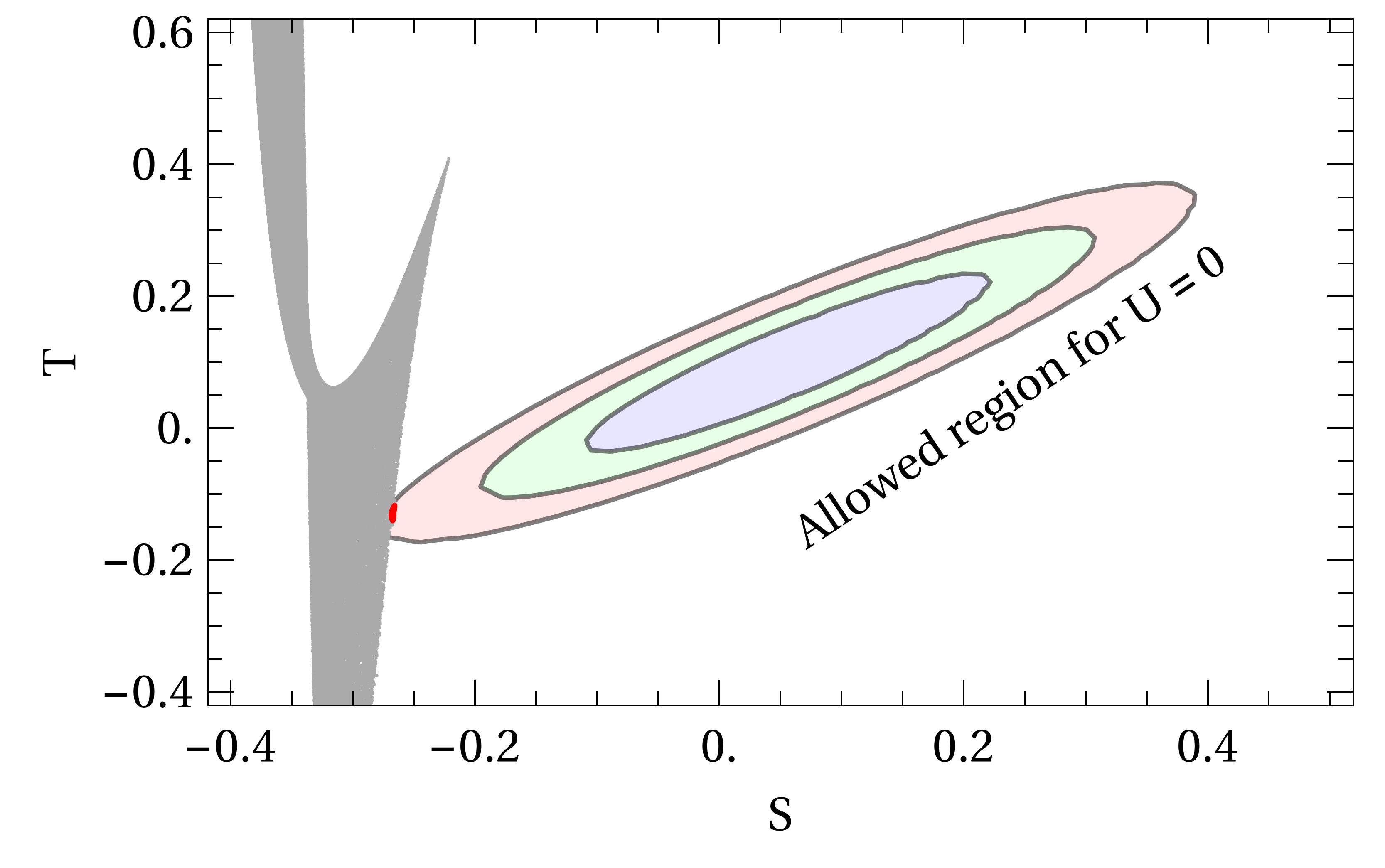}  
  \end{center}
  \caption{Neutrinophilic 2HDM with $\Z_2$ symmetry. The red points
    are allowed by electroweak precision data (oblique parameters) at
    $3\sigma$, while the gray points are ruled out at $3\sigma$ or more 
    in the $S\times T$ plane. No point was found within the $2\sigma$ region.}
  \label{fig:gabriel-nandi-TH}
\end{figure}

{\bf 2HDM with a $\Z_2$ symmetry.} Let us first discuss the results
for the 2HDM with a $\Z_2$ symmetry.  As discussed in
sec.~\ref{subsec:z2}, the model has a very light neutral scalar.  In
fact, we verified that eq.~(\ref{eq:mh}) and the perturbative
unitarity conditions (\ref{eq:unitarity}) require $m_H\lesssim 10
\times v_2$.  Moreover, as the scalar potential parameters $\lambda_i$
and $m_{ij}^2$ can be written in terms of the physical masses and the
vevs, we perform a scan in the physical parameter space, imposing the
following conditions
\begin{align*}
  0.01~{\rm eV} < &m_H < 1~\GeV, \\
  124.85~\GeV < &m_h < 125.33~\GeV, \\
  70~\GeV < &m_{H^\pm}< 1~\TeV,\\
  1~\GeV < &m_A < 1~\TeV,\\
  -\pi/2<&\alpha<\pi/2,\\
  0.01~{\rm eV}<&v_2<1~\MeV.
\end{align*}
The Higgs mass range is taken from the ATLAS$+$CMS measurements combination in
ref.~\cite{Aad:2015zhl}. Note that although $\alpha$ has to be small we did a 
scan over the whole physical range of this parameter, since we wanted to be as general
as possible. By using a logarithmic prior, for convenience, we found that 
indeed $\alpha$ is small.

Using the power of
perturbative unitarity constraints, we found that the
CP-odd and charged scalars are restricted to be below $\sim
600-700~\GeV$. This can be easily understood from eqs.~(\ref{eq:mA})
and (\ref{eq:mHpm}). Since $M^2\propto m_{12}^2=0$ and $\lambda_{4,5}$
cannot be too large, the masses cannot go arbitrarily above the
electroweak vev.

Moreover, the presence of a very light scalar in the spectrum, below
the GeV scale, yields a substantial negative contribution to the $S$
parameter. The impact of the EWPT can be seen in
fig.~(\ref{fig:gabriel-nandi-TH}), where all points scanned were
projected in the $S \times T$ plane and the allowed region by EWPT was
drawn.  Remarkably, only very few points (in red) were found which
provide a viable model, within the 3$\sigma$ allowed region for the
EWPT. 
From our scan, it can be concluded that: the $T$ parameter strongly prefers $m_A
\approx m_{H^\pm}$ or a lighter $H^\pm$ with $m_{H^\pm}\sim 150~\GeV$
together with a $m_A > 300~\GeV$; while the $S$ parameter, although it
depends very mildly on the charged and pseudoscalar masses, exhibits a
slight preference to this latter region. All in all, the values of $S$
are always below $\sim -0.25$, revealing a tension with EWPT always
above the $2.97\sigma$ level.\footnote{To be precise about such strong
  statement, we also included in our analysis the accepted points of
  a second scan centered on the red region, where the charged scalar
  mass range was changed to 150--160~GeV and the pseudoscalar mass
  range was changed to 500--580~GeV, with $10^5$ points.} As an example, we obtained the following 
scalar spectrum, which is allowed at 2.99$\sigma$:
\begin{align*}
m_H&=0.18\ \text{eV},\ m_h=124.9\ \text{GeV},\ m_{H^\pm}=158\ \text{GeV},\\
m_A&=567\ \text{GeV},\ \tan\alpha=-9.3\times 10^{-6},\ \tan\beta=2.3\times 10^{-6}.
\end{align*}
\begin{figure}[t]
  \begin{center}
    \includegraphics[scale=0.6]{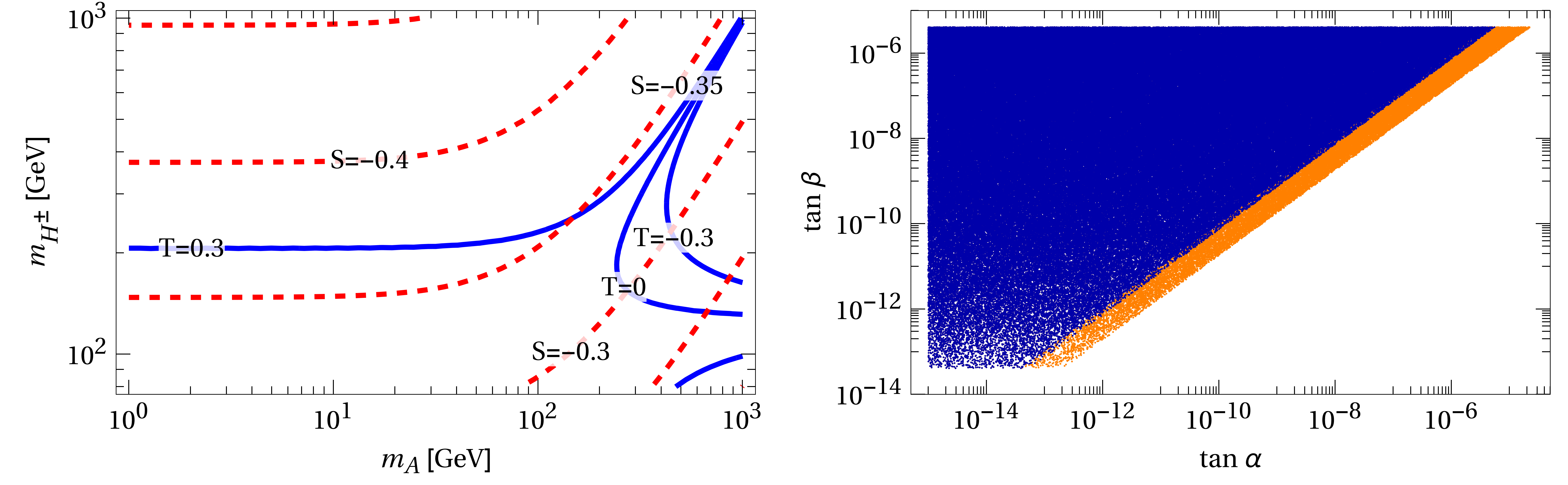}
  \end{center}
  \caption{Neutrinophilic 2HDM with $\Z_2$ symmetry. Left: Predicted
    values for $S$ and $T$ (left) and isolines of $S$ and $T$ values
    as a function of $m_A$ and $m_{H^\pm}$, for $m_H\ll m_Z$. Right:
    $\tan\alpha\times\tan\beta$ plane exclusions obtained using $h$ invisible width.
    Orange points are excluded, while blue points are allowed. }
  \label{fig:gabriel-nandi-TH-2}
\end{figure}
From this analysis, we can conclude that the 2HDM with a $\Z_2$
symmetry is definitely very disfavored by data. It is not even clear
that the region found which is in the $3\sigma$ border of EWPT is
really viable. A closer look into this region of the parameter space
reveals that these points suffer from at least one of the following
worrisome situations: (i) the $e_1$ scattering amplitude, in
eq.~(\ref{eq:unitarity}), is on the verge of violating unitarity, with
at least about $\sim98\%$ of the bound saturated; (ii) the same for
$a_+$ scattering amplitude, with at least $\sim 98\%$ of the bound
saturated; (iii) the stability condition is very fragile, with the
third condition of eq.~(\ref{contrainte}) satisfied with a 
relative difference of less than $\sim 4\times 10^{-4}$; and (iv) the same 
but for the second condition of eq.~(\ref{contrainte}), satisfied 
with a relative difference of less than
$\sim0.05$.  Therefore, given this delicate region of the parameter
space, it would be important to include radiative corrections to see
if the stability and unitarity of the model still holds at one loop.
Notice that by using $4\pi$ as the perturbative unitarity limit
this small region disappears.

A possible way to evade these problems could be
to have a larger $v_2$ so that the mass spectrum, specially $m_H$,
becomes more flexible. Nevertheless, unless $v_2\gtrsim\O(\GeV)$, the
problem does not disappear, strongly disfavoring this minimal model as
an explanation for neutrino masses.

One could now be tempted to include a right-handed neutrino
contribution, dropping the lepton number conservation symmetry of the
model. In fact, as $v_2$ is small, it may be possible to have a low-energy realization of the type I seesaw scenario which leads to
observable sterile neutrino phenomenology, and hopefully could
increase a bit the value of the $S$ parameter to make the model
viable. As the effect on $S$ grows with the mass of the fermions in
the loop, we make a distinction between two regimes: the right-handed
neutrinos can be below or above the GeV scale. In the first, what
happens is that the contribution to the $S$ parameter is suppressed by
the ratio between these small masses and the $Z$ mass and can be
neglected (for instance, the active neutrino contribution to $S$ is
virtually zero). In the second case, although the sterile neutrino
masses might be large, the coupling to the $Z$ is suppressed by the
active--sterile mixing which generically goes as the ratio between the
active to sterile neutrino masses, $m_\nu/m_N$. Therefore the impact
of right-handed neutrinos is never large enough to substantially
change the $S$ parameter\footnote{This fact has also been checked
  numerically using the expressions in ref.~\cite{Akhmedov:2013hec}.}.

For completeness, we also show in the right panel of
figure~(\ref{fig:gabriel-nandi-TH-2}) the impact of the Higgs invisible
width measurement in the $\tan\alpha\times\tan\beta$ plane. Given the
preference for heavier $\S$, we will consider the case where only $h\to
HH$ is present.  From eq.~(\ref{eq:Hhh}), since $m_{12}^2=0$ in this
model, the $g_{hHH}$ coupling can be rewritten in the limit of
small $\beta$ and $\alpha$ as
\begin{equation}
  g_{h H H} \approx -\frac{m_h^2}{v} \dfrac{\sin(2\alpha)}{\sin(2\beta)},
\end{equation}
which can be sizable only if $\alpha \gtrsim \beta$, explaining the
behavior of the excluded region (orange) in
fig.~(\ref{fig:gabriel-nandi-TH-2}).  Since the ratio $\alpha/\beta$
is already constrained by the theoretical limits (see
eq.~(\ref{eq:lambda2})), this constraint turns out to be less
stringent than the others.  As a last comment, the charged scalars
could also have an impact on $h\to\gamma\gamma$. In the small
3$\sigma$ allowed region, the modifications to the diphoton width are
generically between $\pm10\%$, depending on the precise values of
$\lambda_3$. This quartic coupling only affects $m_H$, so it is only
weakly bounded by perturbative unitarity.

Finally, since the smallness of $m_H$ causes the tension with EWPT, one may wonder what is the impact of loop corrections on the scalar spectrum of this model. Generically, in a 2HDM with $\mathbb{Z}_2$ symmetry, the charged and CP-odd mass matrices are not modified by one-loop corrections. The CP-even matrix receives radiative corrections of the form \cite{Lee:2012jn}
\begin{equation}
M_\rho = 
\begin{pmatrix}
\lambda_1 v_1^2 & \lambda_{345} v_1 v_2\\ 
\lambda_{345} v_1 v_2 &  \lambda_2 v_2^2 
\end{pmatrix}+\frac{1}{64\pi^2}
\begin{pmatrix}
\Delta m_{11}^2 v_1^2 & \Delta m_{12}^2 v_1 v_2 \\
\Delta m_{12}^2 v_1 v_2  & \Delta m_{22}^2 v_2^2 
\end{pmatrix},
\end{equation}
where the second term comes from the one-loop effective potential. Since $\Delta m_{ij}^2$ are solely functions of masses and quartic couplings, the dependence of the CP-even mass matrix on $v_{1,2}$ is preserved at one-loop level, implying a small value for $m_H$ if $v_2$ is small. We have checked by explicit calculations that the corrections to $m_H$ are at the most a factor $100$, which is still insufficient to solve the problem with the $S$ parameter.

\begin{figure}[t]
  \begin{center}
    \includegraphics[scale=0.6]{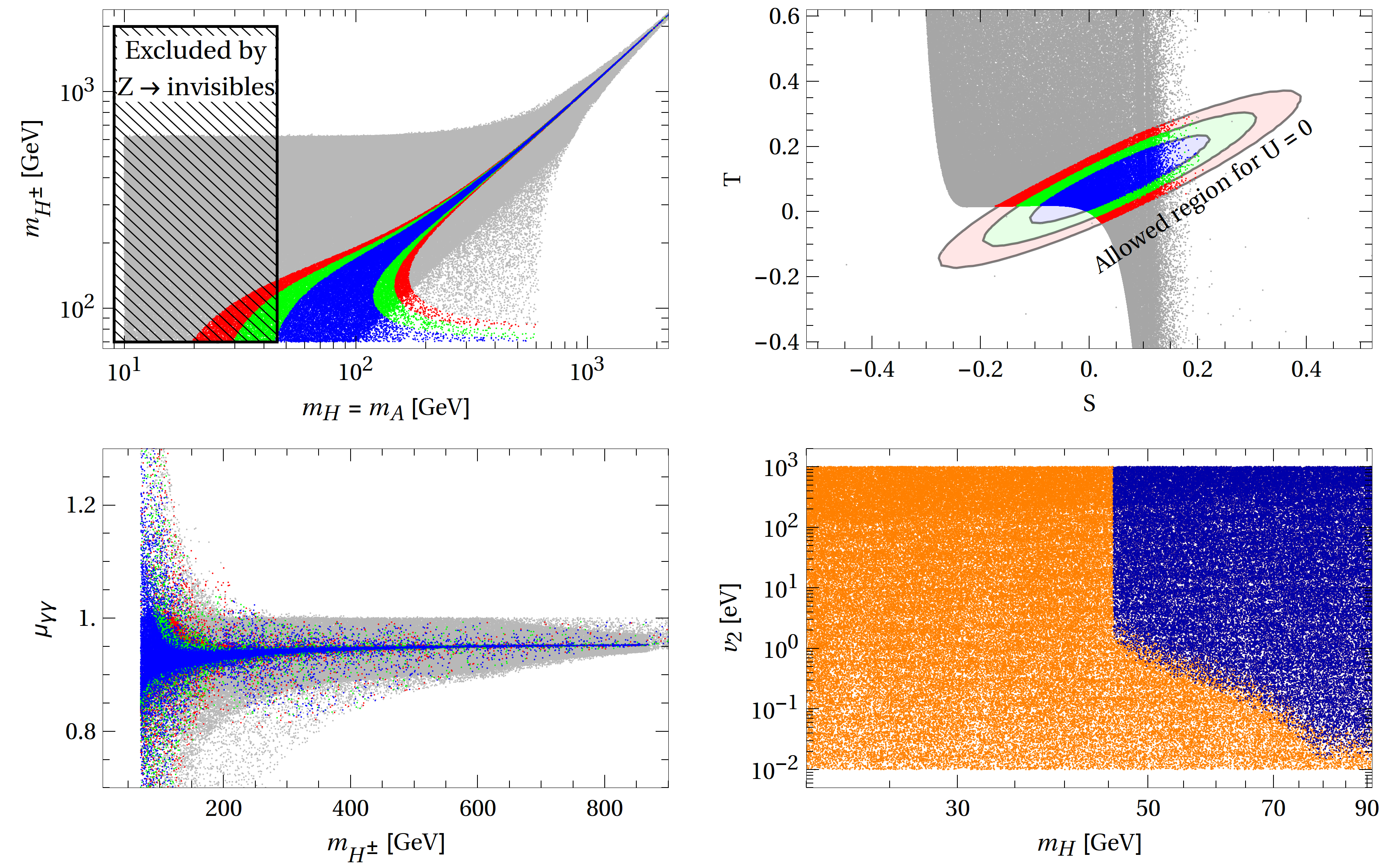}
  \end{center}
  \caption{Neutrinophilic 2HDM with softly broken global $U(1)$
    symmetry. The blue, green and red points are allowed by EWPT at
    $1\sigma$, $2\sigma$, and $3\sigma$, respectively, while the gray
    points are ruled out at $3\sigma$. Top left: parameter space in
    the plane $m_H \times m_{H^\pm}$ which satisfy perturbativity, 
    unitarity and stability constraints. Top right: projection of
    these points in the $S\times T$ plane. Bottom left:
    $h\to\gamma\gamma$ signal strength as a function of
    $m_{H^\pm}$. Bottom right: region in the $m_H\times v_2$ plane
    that is excluded by the Z invisible width (orange
    points).}
  \label{fig:davidson-logan-TH}
\end{figure}

{\bf 2HDM with a global $U(1)$ symmetry.}  We now focus on the
phenomenology of the softly broken $U(1)$ model. A non-zero $m_{12}^2$
term allows for heavier $H$, presenting a major change in the
phenomenology with respect to the previous model. Without the
requirement of a light scalar, we enlarge the scanned region
accordingly.  The absence of the $\lambda_5$ quartic coupling makes
the pseudoscalar degenerate in mass with $H$ (to first order in
$v_2$). Therefore we perform an initial scan of the spectrum parameter
space, this time in the region
\begin{align*}
  10~\GeV < &m_H < 1~\TeV, \\
  124.85~\GeV < &m_h < 125.33~\GeV, \\
  70~\GeV < &m_{H^\pm}< 1~\TeV,\\
  m_A &= m_H,\\
  -\pi/2<&\alpha<\pi/2,\\
  0.01~{\rm eV}<&v_2<1~\MeV,
\end{align*}
as well as a second scan with $m_{H^\pm}$ and $m_A$ heavier then 1 TeV
and almost degenerate.  We follow the same procedure as before,
showing only the points allowed by perturbative unitarity and
stability constraints.  The results are presented in
fig.~(\ref{fig:davidson-logan-TH}). 

In contrast to the previous case, due to the possibility of obtaining
a heavier $H$ in the mass spectrum, there is a region of the parameter
space of this model which passes the electroweak precision tests and
theoretical constraints. The behavior of the $T$ parameter is similar
to the previous scenario: either the mass splitting between $A$ and
$H^{\pm}$ is at most $\sim 80~\GeV$, or the charged scalar is around
100~GeV while $m_H=m_A>150~\GeV$, with negative values of $T$ for
larger $m_{H^\pm}$. This explains the strong correlation on the
allowed region in the upper left panel of
fig.~(\ref{fig:davidson-logan-TH}).
We also present the projection of these points in the $S\times T$
plane in the upper right panel of fig.~(\ref{fig:davidson-logan-TH}).
For the $\alpha$ and $\beta$ parameters we find that the 
allowed region is $\tan\beta\lesssim 10^{-6}$ and $\alpha\lesssim 5\beta$.

As discussed in the previous sections, this model can also accommodate
a pair of neutral scalars ($\S=H,A$) satisfying $m_\S< m_h/2$ if
$m_{12}^2$ is small enough. In this case, the constraints coming from
the Higgs invisible decays are similar to those described for the
model with a $\mathbb{Z}_2$ symmetry and turn out to be relatively
weak. On other hand, the $Z$ invisible width can provide  valuable
constraints when the channel $Z\to \S\nu\nu$ is open. To perform this
analysis we scan over the oscillation parameters, imposing the 
perturbativity condition $\Gamma_\S<m_\S/2$. We show on the bottom 
right panel of figure~(\ref{fig:davidson-logan-TH}) the excluded 
region (orange points) under these assumptions in the $m_H\times v_2$ 
plane. The region $m_\S<m_Z/2$ is completely excluded, because in 
this case we integrate over the poles of the off-shell scalars 
in $Z\to H (A^\ast\to \nu\bar{\nu})$ and $Z\to A (H^\ast\to \nu\bar{\nu})$, 
enhancing the decay rate by orders of magnitude. 

For a heavy enough $H^{\pm}$, as can be seen in the lower left panel of
fig.~(\ref{fig:davidson-logan-TH}), the $h\to\gamma\gamma$ signal
strength is diminished by about $\sim 5\%$. We can understand this non
decoupling feature by noticing that the $h\,H^+ H^-$ coupling is
$-i\lambda_3v$, which in turn has a correlation with $m_{H^\pm}$,
specially in the larger mass region.  This can be understood by
noticing that, in eq.~(\ref{lambda3}), for large $m_{H^\pm}$, we have
\begin{equation}
  \lambda_3\approx\left(1-\dfrac{\sin 2\alpha}{\sin 2\beta}\right)\dfrac{m_{H^\pm}^2}{v^2}.
\end{equation}
Typically, $\alpha \lesssim 5 \beta$, which corresponds to a strong
correlation between $\lambda_3$ and $m_{H^\pm}$, and this is the
denser region around $\mu_{\gamma \gamma}=0.95$. However this is not always the case,
and the correlation is lost when the ratio of sines is closer to 1,
now corresponding to the sparser points with a much weaker
correlation. Nevertheless, we see that for a heavy enough charged
scalar, the contribution to the Higgs diphoton width is always
negative. 

One could ask if it is also possible to have a Majorana mass term,
since the $U(1)$ symmetry is softly broken. First, as pointed out in 
refs.~\cite{Akhmedov:2013hec, Fernandez-Martinez:2015hxa}, the impact of heavy right-handed neutrinos 
via loop effects on electroweak precision observables is very small. 
Therefore, there is no significant interplay between this and the scalar 
sector of the model, and thus the phenomenology studied here would be 
essentially unchanged. On the other
hand, if we consider an UV completion that simultaneously breaks the symmetry
and originates a Majorana mass term, we find that such scenario is non-minimal, i.e., at least two new fields have to be included.

\subsection*{Comments on non-minimal models}
\label{sec:other}

Due to the large number of possible variants, performing exhaustive
analises of non-minimial models is unpractical and well beyond the purpose of this paper. Nonetheless, we may
glimpse the phenomenology of some representative cases.

A neutrinophilic 2HDM with a softly broken $\Z_2$ symmetry would
surely be allowed by data, in contrast to the spontaneously broken
$\Z_2$ scenario. Such model would be more general than the two models
considered here, as it would span a non-zero value of both $m_{12}$
and $\lambda_5$. As can be seen from eqs.~(\ref{eq:mh}) and
(\ref{eq:mA}),  the simultaneous presence of these terms in the 
scalar potential lifts the degeneracy between the neutral scalars 
$A$ and $H$. In fact, we have checked that the allowed region in the plane $m_A\times m_{H^\pm}$ 
is very similar to the one exhibited in
fig.~(\ref{fig:davidson-logan-TH}) (top left panel), except for the
fact that the $T$ parameter now implies a correlation only between
$m_{H^\pm}$ and $m_A$. If the neutral scalar $H$ decays dominantly to
neutrinos, as it is likely to happen, it would be very difficult to
probe it by resonant production at colliders.

Another way of evading our limits would be to enlarge the particle spectrum of the spontaneously
broken $\Z_2$ model (or generically any $\Z_N$), for instance, by
adding a scalar singlet $S$, doublet $\Phi_3$ or triplet $\Delta$, all
charged under the new symmetry. In the singlet case, a triple or
quartic term $S \Phi_1^\dagger \Phi_2$ or $S^2 \Phi_1^\dagger \Phi_2$
could be present in the potential for a judicious choice of charges.
After the singlet acquires a vev, this term would play a role similar
to the $m_{12}$ soft breaking term allowing for larger values of $m_H$. However, the quartic
$S^\dagger S\Phi_1^\dagger\Phi_1$, always present, would induce a
Higgs-singlet mixing. This would diminish all Higgs couplings to
fermions and gauge bosons by a factor $\sin \theta$ where $\theta$ is
the corresponding mixing angle. The mixing is constrained by Higgs
production cross section measurements to be $\sin^2 \theta \lesssim
0.2$~\cite{ATLAS-CONF-2015-044}. In the case of adding a third doublet, the triple coupling is impossible, but a quartic one could be present. Last, in the case of a scalar triplet, although a triple coupling would be
possible, a large triplet vev could irrevocably disturb electroweak
precision tests, especially the $T$ parameter. 

\section{Conclusion}\label{sec:conc}
We performed an analysis of the minimal neutrinophilic two-Higgs-doublet
models which can accommodate neutrino masses by means of the tiny vev of
the additional Higgs doublet. The models studied here differ among
themselves by the symmetry that forbids the couplings between 
neutrinos and the scalar which gets the electroweak scale vev. 
The cases studied here span a discrete $\Z_2$ and a softly broken global $U(1)$ 
symmetry.

The bounds considered come both from theory and experiment. The
unitarity perturbative requirement at tree-level strongly constrains
the scalar mass spectrum of these models, either by the presence of a
very light neutral scalar ($m_H \sim v_2$), in the $\Z_2$ model, or
with a degeneracy between the scalar and pseudoscalar particle masses
($m_H =m_A$), in the global $U(1)$ scenario.

If there is no additional particle content, the $\Z_2$ symmetry model
was found to be in severe tension with the electroweak precision
tests, due to the very light neutral scalar, which generates a large
negative contribution to the $S$ parameter. The inclusion of a
Majorana mass term for the right-handed neutrinos, providing a low
scale realization of the seesaw type I mechanism, does not save the
model, as the right-handed neutrino contribution to the $S$ parameter
is always negligible. Therefore, we conclude that \emph{the
  neutrinophilic 2HDM with a spontaneously broken $\Z_2$ symmetry is
  strongly disfavored by data}.
  
The analysis of the model with an explicit broken global $U(1)$
symmetry reveals a region of the parameter space which is allowed by
all bounds considered. Due to the set of constraints and the
symmetries of the model itself, the spectrum is quite limited.
The $U(1)$ symmetry predicts that the neutrinophilic scalar is
degenerate in mass with the pseudoscalar, $m_H=m_A$.  Besides, the
electroweak precision tests play a very important role, specially the
$T$ parameter which is sensitive to the absolute mass splitting of the
pseudoscalar and the charged scalars, limiting it to be at most
$\sim80~\GeV$.  Therefore, an important consequence of the theoretical
and experimental constraints is that, if the new scalars are above
$\sim 400~\GeV$, all these particles should have very similar masses.
Moreover, the $Z$ invisible width excludes the region $m_H=m_A<m_Z/2$. Besides, the
$h\to\gamma\gamma$ branching fraction might be modified by about
$\pm 30\%$ for $m_{H^\pm}<200~\GeV$, while for heavier $H^\pm$,
above 500~GeV, this ratio can be atmost 1 or lower by $5\%$.  
Finally, we stress that this model can be well within the reach of LHC 13~TeV, by probing
the $h\to\gamma\gamma$ branching fraction of by direct pair production
of the charged scalars, if they are below $\O(300~\GeV)$.

\begin{acknowledgments}
It is a pleasure to acknowledge stimulating discussions with Enrico
Bertuzzo. This work was supported by Funda\c{c}\~ao de Amparo \`a
Pesquisa do Estado de S\~ao Paulo (FAPESP) and Conselho Nacional de
Ci\^encia e Tecnologia (CNPq). RZF would like to thank the hospitality
of the LPT-Orsay, where part of this work was completed. PM
acknowledges partial support from the European Union FP7 ITN
INVISIBLES (Marie Curie Actions, PITN-GA-2011-289442), and from the
Spanish MINECO's ``Centro de Excelencia Severo Ochoa'' Programme under
grant SEV-2012-0249.
\end{acknowledgments}

\clearpage

\appendix
\section{Tree-level Relations for the Quartic Couplings}
\label{ap:quartic}

The quartic couplings can be expressed in terms of the physical masses, vevs and mixing angles as:~\cite{Kanemura:2004mg}
\begin{align}
\label{lambda1}
\lambda_1  &= \frac{1}{v^2}\left(-\tan^2\beta M^2+\frac{\sin^2\alpha}{\cos^2\beta} m_H^2+\frac{\cos^2\alpha}{\cos^2\beta}m_h^2\right), \\
\label{lambda2}
\lambda_2  &=\frac{1}{v^2}\left(-\cot^2\beta M^2+\frac{\cos^2\alpha}{\sin^2\beta} m_H^2+\frac{\sin^2\alpha}{\sin^2\beta}m_h^2\right), \\
\label{lambda3}
\lambda_3 &=\frac{1}{v^2}\left(-M^2+2 m_{H^\pm}^2+\dfrac{\sin(2\alpha)}{\sin(2\beta)}(m_h^2-m_H^2)\right),\\
\label{lambda4}
\lambda_4 &= \frac{1}{v^2} \left(M^2+m_A^2-2 m_{H^\pm}^2\right) \\
\label{lambda5}
\lambda_5 &= \frac{1}{v^2} \left(M^2- m_A^2\right),
\end{align}
where $M^2\equiv\dfrac{m_{12}^2}{\sin\beta \cos\beta}$. Inversely, we have 
\begin{align}
m_h^2 &= M^2 \sin^2(\alpha-\beta)\nonumber\\ \label{eq:mH}
&\quad +\left(\lambda_1 \cos^2 \alpha \cos^2\beta+\lambda_2 \sin^2 \alpha \sin^2\beta+\frac{\lambda_{345}}{2}\sin 2\alpha \sin 2\beta\right)v^2, \\ 
m_H^2 &= M^2 \cos^2(\alpha-\beta)\nonumber\\ \label{eq:mh}
&\quad+\left(\lambda_1 \sin^2 \alpha \cos^2\beta+\lambda_2 \cos^2 \alpha \sin^2\beta-\frac{\lambda_{345}}{2}\sin 2\alpha \sin 2\beta\right)v^2, \\ \label{eq:mA}
m_A^2 &= M^2 - \lambda_5 v^2, \\ \label{eq:mHpm}
m_{H^\pm}^2 &= M^2 - \frac{\lambda_{45}}{2} v^2,
\end{align}
where $\lambda_{45}=\lambda_4+\lambda_5$ and $\lambda_{345}=\lambda_3+\lambda_4+\lambda_5$.

\clearpage
\bibliographystyle{JHEP}
\bibliography{./neutrino-2hdm}

\end{document}